\documentclass[]{spie} 
\usepackage[]{graphicx}

\title{Science with the Keck Interferometer ASTRA Program} 

\author{J. A. Eisner\supit{a}, R. Akeson\supit{b}, M. Colavita\supit{c}, A. Ghez\supit{d},
 J. Graham\supit{e}, L. Hillenbrand\supit{f}, R. Millan-Gabet\supit{b}, 
 J.D. Monnier\supit{g}, J.-U. Pott\supit{h}, S. Ragland\supit{i},
P. Wizinowich\supit{i},  J. Woillez\supit{i}
 \\
\vspace{0.1in}
\supit{a}Steward Observatory, University of Arizona, Tucson, AZ 85721 \\
\supit{b}NASA Exoplanet Science Institute, Caltech, Pasadena, CA 91125 \\
\supit{c}Jet Propulsion Laboratory, Caltech, Pasadena, CA 91109 \\
\supit{d}Astronomy Department, UCLA, Los Angeles, CA 90095 \\
\supit{e}Astronomy Department, UC Berkeley, Berkeley, CA 94720 \\
\supit{f}Astrophysics Department, Caltech, Pasadena, CA 91125 \\
\supit{g}Astronomy Department, University of Michigan, Ann Arbor, MI 48109 \\
\supit{h}Max-Planck-Institut f\"{u}r Astronomie, K\"{o}nigstuhl 17, D-69117 Heidelberg, Germany \\
\supit{i}W.M. Keck Observatory, Kamuela, HI 96743 \\
}

\begin{document} 
  \maketitle 

\begin{abstract}
The ASTrometric and phase-Referenced Astronomy (ASTRA) project will
provide phase referencing and astrometric observations at the Keck
Interferometer, leading to enhanced sensitivity and the ability to
monitor orbits at an accuracy level of 30-100 microarcseconds.  Here
we discuss recent scientific results from ASTRA, and describe new
scientific programs that will begin in 2010--2011.  We begin with results
from the ``self phase referencing" (SPR) mode of ASTRA, which uses
continuum light to correct atmospheric phase variations and produce a
phase-stabilized channel for spectroscopy.  We have observed a number
of protoplanetary disks using SPR and a grism providing a spectral
dispersion of $\sim 2000$.  In our data we spatially resolve emission
from dust as well as gas. Hydrogen line emission is spectrally
resolved, allowing differential phase measurements across the emission
line that constrain the relative centroids of different velocity
components at the 10 microarcsecond level.  In the upcoming year, we
will begin dual-field phase referencing (DFPR) measurements of the
Galactic Center and a number of exoplanet systems.  These observations
will, in part, serve as precursors to astrometric monitoring of
stellar orbits in the Galactic Center and stellar wobbles of exoplanet
host stars.  We describe the design of several scientific
investigations capitalizing on the upcoming phase-referencing and
astrometric capabilities of ASTRA.
\end{abstract}



\section{INTRODUCTION}
\label{sec:intro}  
ASTrometric and phase-Referenced Astronomy (ASTRA) is a program of upgrades to the Keck 
Interferometer funded by an NSF MRI grant.  ASTRA combines the power of laser guide star 
adaptive optics (LGS-AO) on the twin 10-m Keck apertures with
dual-field phase referencing (DFPR) to enable interferometric
observations of stars fainter than $m_K = 14$.  Precise measurement
and monitoring of the baseline separating the two Keck apertures allows astrometric observations
with an accuracy of $< 50$ $\mu$as.  These capabilities enable a broad range of new science,
and we describe ongoing and upcoming investigations below.

\section{ASTRA Capabilities}
\begin{figure}
  \begin{minipage}{9.0cm}
    \centerline{\includegraphics[clip=true,scale=1.3]{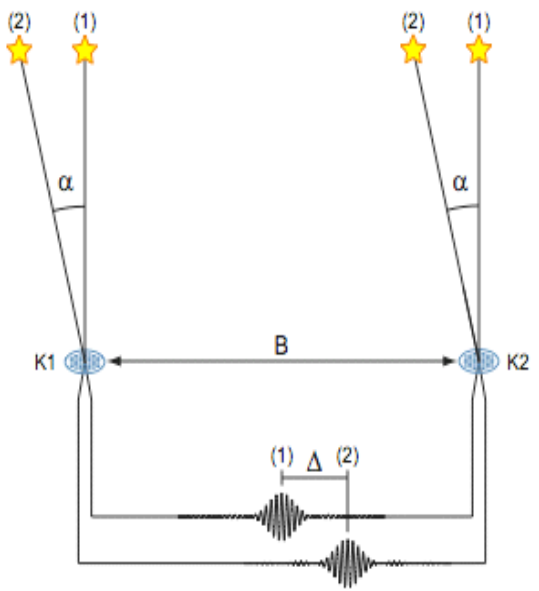}}
  \end{minipage} \hfill
  \begin{minipage}{7.0cm}
\caption{ \label{fig:dualstar} 
A sketch illustrating the principle of interferometric phase-referencing. 
The sketch depicts a ``dual-star'' setup where light from a bright reference
star is used to measure atmospheric phase as a function of time.  
Light from a fainter science target is sent down a separate beam-train
that is phase-stabilized using the information from the reference channel.
This setup allows longer integrations, and hence greater sensitivity, for 
the science target.  SPR uses a simplified setup, where
the reference and science targets are actually the same object.  Continuum
light from a source is used to correct the atmosphere, and spectrally-dispersed
light is sent down a phase-stabilized beam-train.
}
\end{minipage}
   \end{figure}

Dual-field phase referencing is analogous to natural guide star AO, in that it uses light
from a bright reference star to correct the atmosphere and allow longer integrations.
ASTRA uses the light from the reference star to measure the atmospheric phase, and a
servo loop then removes these phase motions from a second beam train.  A fainter source
can be observed with longer integration times using this phase-stabilized second beam 
train.  A sketch of the dual-star setup used by ASTRA is shown in Figure \ref{fig:dualstar}.

As a precursor to DFPR, we implemented a simple setup called ``self phase-referencing'' (SPR).
Instead of a faint science target and a bright reference star, we split the light from a single, bright 
star into a continuum channel and a spectrally dispersed channel.  Since all of the light is on-axis 
in this case, we can split the light near to the beam combination optics, minimizing the non-common
path between the two beamtrains.  Spectrally dispersed fringes can be observed in the phase-stabilized
beamtrain with long integration times, allowing high signal-to-noise spectroscopic observations.
We have implemented a grism in the secondary camera with a spectral resolution of $R \approx 2000$
to capitalize on this capability.

Once DFPR is commissioned, the next step is astrometry.  DFPR measures
the delays of stars in both fields.  If the baseline separation
between the two telescopes, and the instrumental path traversed by
light from the two sources, is known then the measured delays can be
converted into a separation on the sky.  ASTRA aims to deliver 30--100
$\mu$as astrometric measurement accuracy.

Further description of the ASTRA SPR, DFPR, and astrometric modes can
be found in papers by Woillez et al. and Ragland et al. in this
volume.

\section{SPECTROSCOPY OF PROTOPLANETARY DISKS}
\label{sec:spr}

\begin{figure}[bp]
  \begin{center}
  \includegraphics[scale=0.9,clip=true]{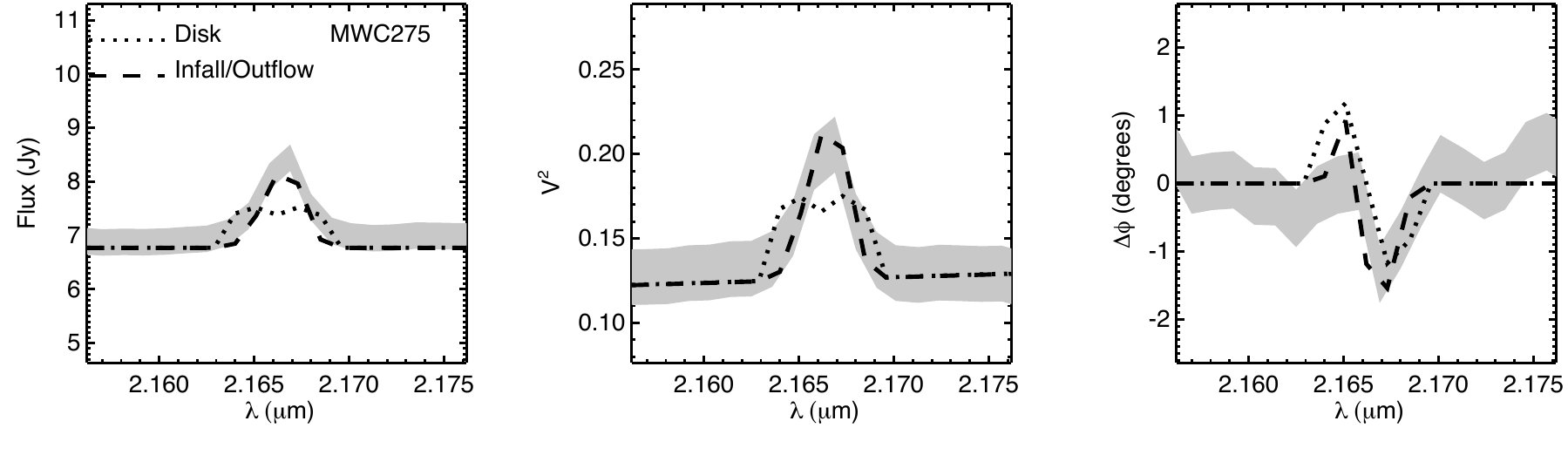}
   \end{center}
   \caption{ \label{fig:brg} 
Example of SPR data from KI/ASTRA, with predictions of models over-plotted.  The gray regions indicate the 
1-$\sigma$ error intervals on the data, after removal of the stellar component.  The dotted curves show the
predictions of a Keplerian disk model fitted to the data, and the dashed curves show the predictions of a
bipolar infall/outflow model.  Figure adapted from Eisner et al. 2010.}
   \end{figure} 

Recently, spectrally dispersed interferometry has enabled probes of
the gas in protoplanetary disks
\cite{EISNER+07a,EISNER07,MALBET+07,TATULLI+07,TATULLI+08,ISELLA+08,
  KRAUS+08,EISNER+09}.  Many of these studies focused on Br$\gamma$
emission, which traces hydrogen gas at $\sim 10,000$ K, cascading down
to the ground state from highly excited electronic states after recent
recombination.  Most of these studies found Br$\gamma$
emission to be more compactly distributed than continuum emission
\cite{EISNER07,KRAUS+08,EISNER+09}, although more extended
distributions were seen in a few cases \cite{MALBET+07,TATULLI+07}. 
However, the Br$\gamma$ emission was only spectrally resolved for a
few of these objects \cite{KRAUS+08}, all of which are bright Herbig
Ae/Be stars.

We have used the SPR/grism mode of ASTRA to spectrally and spatially
resolve Br$\gamma$ emission from a sample of protoplanetary disks
around stars spanning a broad spectral type range.  While 
our observations covered approximately the entire $K$-band, most of
our early work focused on the spectral region around the Br$\gamma$
line of hydrogen at 2.1662 $\mu$m.  Measured fluxes, visibilities, and
differential phases versus wavelength constrain the spatial
distribution and kinematics of the hydrogen gas within 0.1 AU of the
central stars in these systems.  

The angular resolution of these observations corresponds to
stellocentric radii of $\sim 0.1$ AU at the distances to our young
star targets.  However, we used the differential phases observed
across the resolved Br$\gamma$ lines to measure spectro-astrometric
offsets of various velocity components.  These spatial offsets can be
measured with an uncertainty of $<0.01$ AU.

We found that in many sources, the inferred size of the Br$\gamma$
emission was $<0.01$ AU, implying an origin in accretion columns very
close to the central star.  For other objects, the average size of the
Br$\gamma$ emission was somewhat larger, $\sim 0.1$ AU.  However,
analysis of the spectrally resolved data for these objects indicated
at least some emission on scales of 0.01 AU or smaller.  Thus, we
suggest that Br$\gamma$ emission generally traces accretion, although
some Br$\gamma$ emission may also arise from the inner regions of
outflows in some cases.  This work is described in detail in Eisner et
al. (2010).  An example  of data, with model fits, is shown in
Figure \ref{fig:brg}.

\begin{figure}
  \begin{center}
 \includegraphics[scale=1.4,clip=true]{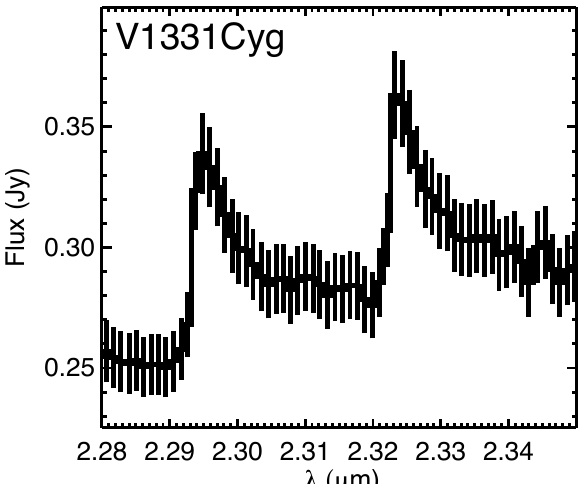}
  \includegraphics[scale=1.36,clip=true]{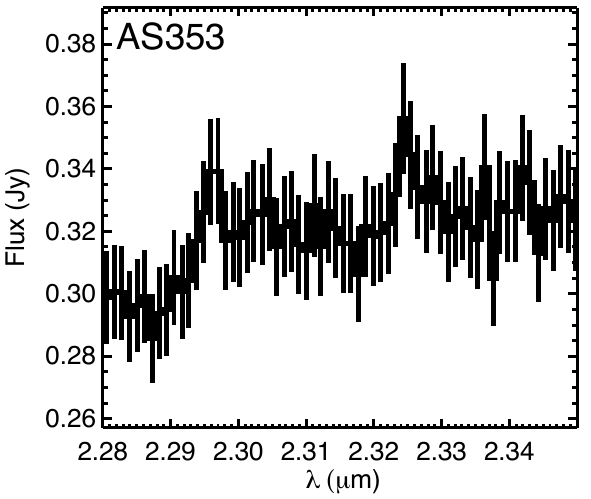}
   \end{center}
   \caption{ \label{fig:co} 
Example spectra of CO overtone bandhead emission in protoplanetary disks, taken with KI/ASTRA.
These objects are fainter than the one shown in Figure \ref{fig:brg}, and so we were unable to observe $V^2$ and
differential phase signatures with high signal-to-noise across these spectral features.  In the future, we plan to 
employ longer integrations to obtain such measurements.}
   \end{figure}

In the future, we will extend such observations to fainter
sources--using longer integrations--and to other spectral regions.  In
particular, we will probe regions where molecular emission is
expected, and map the distribution of molecular material
in potentially planet-forming regions (Figure \ref{fig:co}).


\section{SPATIALLY RESOLVED OBSERVATIONS OF ACTIVE GALACTIC NUCLEI}
\label{sec:agn}
Optical and infrared interferometry has historically been limited to
bright targets, and so had little relevance for extragalactic
astronomy.  The enhanced sensitivity enabled by ASTRA's DFPR mode 
changes this.  To date, only the brightest AGNs have been 
observed with interferometers (KI and VLTI). These observations seem
to be consistent with the AGN unification scheme where a molecular
torus explains both type 1 and 2 phenomenology. DFPR will
enlarge the sample of AGNs for studies of the inner edges of
molecular tori.  This will help elucidate the geometrical and
physical nature of the nuclear obscuration, and the geometrical
differences between type 1 and 2 AGNs.  

DFPR observations of AGN can also use the grism described above to
disperse the AGN light in the phase-stabilized channel.  The $R\sim
2000$ of the grism is sufficient to spectrally resolve the broad line
regions in many AGN, and ASTRA observations can thus provide
spectro-astrometric measurements.  With $\sim 10$ $\mu$as
uncertainties on spectro-astrometry with ASTRA (already demonstrated
with SPR\cite{POTT+10,EISNER+10}), spatial offsets across the
Br$\gamma$ line in AGN at a distance of 10 Mpc can be detected at the
level of 100 AU.  This is smaller than even the most compact BLR size
inferred from reverberation mapping \cite{BENTZ+06}.
ASTRA can measure BLR sizes, and hence directly constrain the masses of the
central black holes.

\section{ASTROMETRIC OBSERVATIONS OF EXOPLANETS}
\label{sec:dfpr}

\begin{figure}
  \begin{minipage}{11.0cm}
    \centerline{\includegraphics[clip=true,scale=0.9]{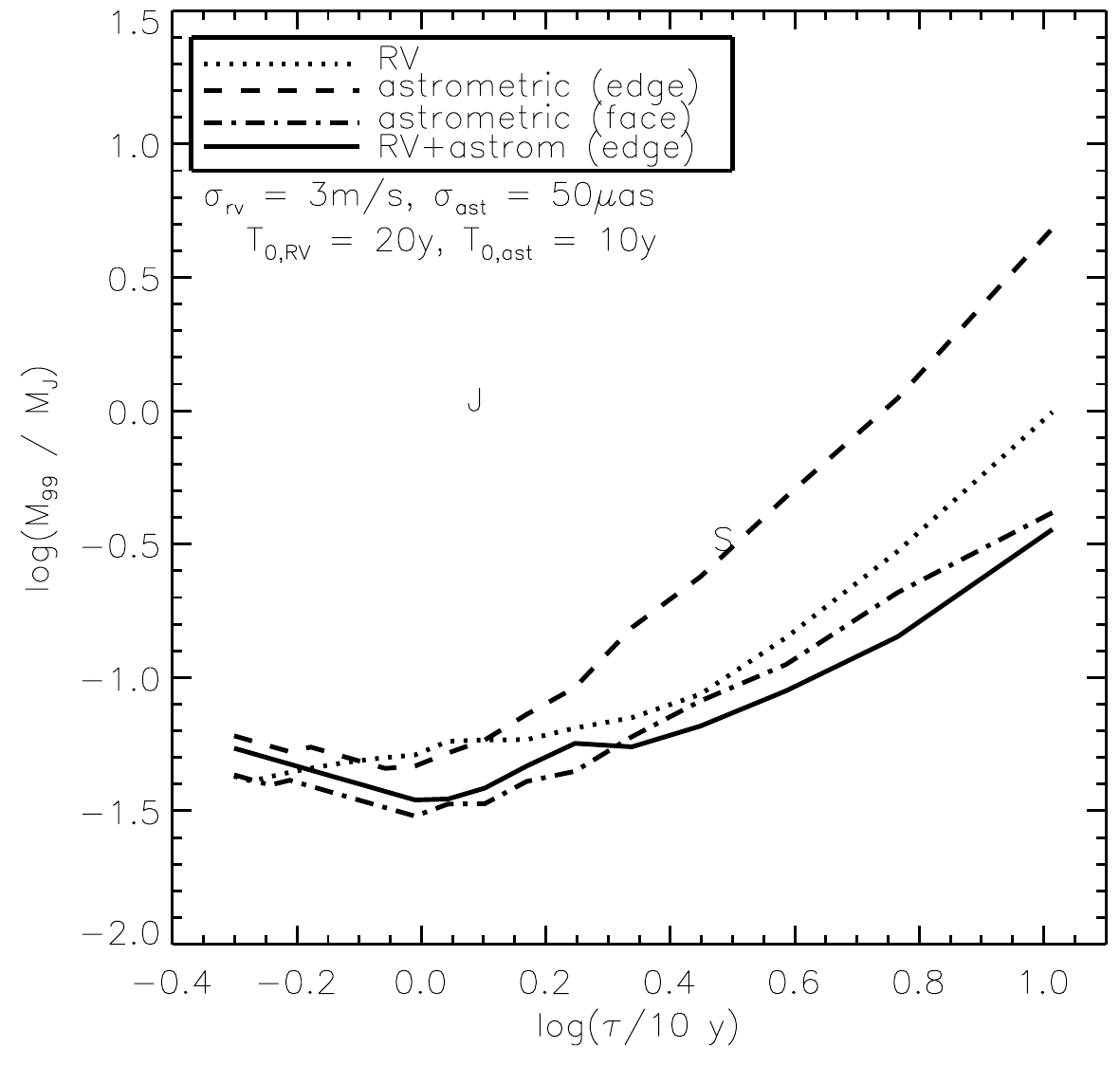}}
  \end{minipage} \hfill
  \begin{minipage}{6.0cm}
\caption{\label{fig:planets} 
     Mass sensitivity (99\% confidence limits) of astrometric,
radial velocity, or combined RV+astrometry observations of a star at
10 pc.  
``J" and ``S'' indicate the locations of Jupiter and Saturn in this
plot.  We assume a 20 year RV survey and a 10 year astrometric
survey, and measurement precisions of  3 m s$^{-1}$
for the RV data and 50 $\mu$as for the astrometry.  In these simulations,
measurements are taken once a month.  These parameters
are chosen to represent current RV surveys and additional data from
ASTRA.  Edge-on systems
only allow 1D astrometry, and so the ``edge''-labeled curve is the
worst-case.  However, even in this case, combining RV and astrometry
data allows much better sensitivity than RV alone.  Astrometry has a
substantial advantage for planets with periods of a few
years.
\vspace{0.9in}
}
\end{minipage}
\end{figure}

The DFPR mode, in conjunction with astrometry, can be used to measure the wobble of stars induced by
orbiting planets.  The center-of-mass of a planet/star system is constant, and so a planet induces a star to
orbit about the center of mass with an amplitude of 
\begin{equation}
\theta_{\rm ast} = 95 \: {\rm \mu as} \:
\left(\frac{M_{\rm planet}}{1 \: {\rm M_{Jup}}}\right)
\left(\frac{d}{10 \: {\rm pc}}\right)^{-1} \left(\frac{\tau}{1 \: {\rm
      yr}}\right)^{2/3} \left(\frac{M_{\ast}}{\rm M_{\odot}}\right)^{-2/3}.
\label{eq:astrom}
\end{equation}
Here, $M_{\rm planet}$ is the planet mass, $d$ is the distance to the
system, $\tau$ is the orbital period of the planet, and $M_{\ast}$ is
the stellar mass.    In addition, astrometric measurements will be
affected by linear proper motion and parallax of the host star.  This
lowers the effective sensitivity to planets in cases where the orbital
period is comparable to or longer than the duration of the astrometric
survey \cite{EK01b}.

With measurements sampling a complete orbit with good phase
coverage, 50 $\mu$as accuracy is sufficient to detect 
Jupiter-mass (or lower-mass) planets around nearby stars (Figure \ref{fig:planets}).  Indeed,
this level of precision is sufficient to detect planets that have gone
undetected by radial velocity surveys (in some cases by combining astrometric and
radial velocity data\cite{EK02}).

Probably the most straightforward application of astrometry is to determine orbital inclinations of
exoplanets already known from radial velocity monitoring surveys.  With inclination determined, masses 
can also be determined without a $\sin i$ ambiguity.  For multiple planet systems, the inclinations are 
also important for understanding the stability and provenance of the planets.

Astrometry also has the potential to detect planets that are inaccessible to radial velocity observations.
For example, A stars show few spectral lines (at least while on the main-sequence), and so are poor targets
for radial velocity surveys.  They are, however, fine targets for astrometry.  Astrometric monitoring of nearby A stars
with KI/ASTRA has the potential to detect massive planets with orbits like that of Jupiter.

A final case that we mention here is planets around pre-main-sequence
stars.  Starspots can cause significant RV variations at optical
wavelengths \cite{DESORT+07}.  In the near-IR, the flux contrast between
the stellar photosphere and the spot is considerably lower, and so
starspot noise is less of a problem for near-IR measurements.  Furthermore,
astrometric amplitudes are typically
small for short-period signals \cite{EK02}, and so starspot
noise with periods near the stellar rotation period are less
significant \cite{MAKAROV+09}.  ASTRA astrometry is thus well-suited
to detecting planets around pre-main-sequence stars.

\section{``IMAGING'' OF THE GALACTIC CENTER}
\label{sec:gcim}

\begin{figure}
  \begin{minipage}{11.0cm}
    \centerline{\includegraphics[clip=true,scale=0.6]{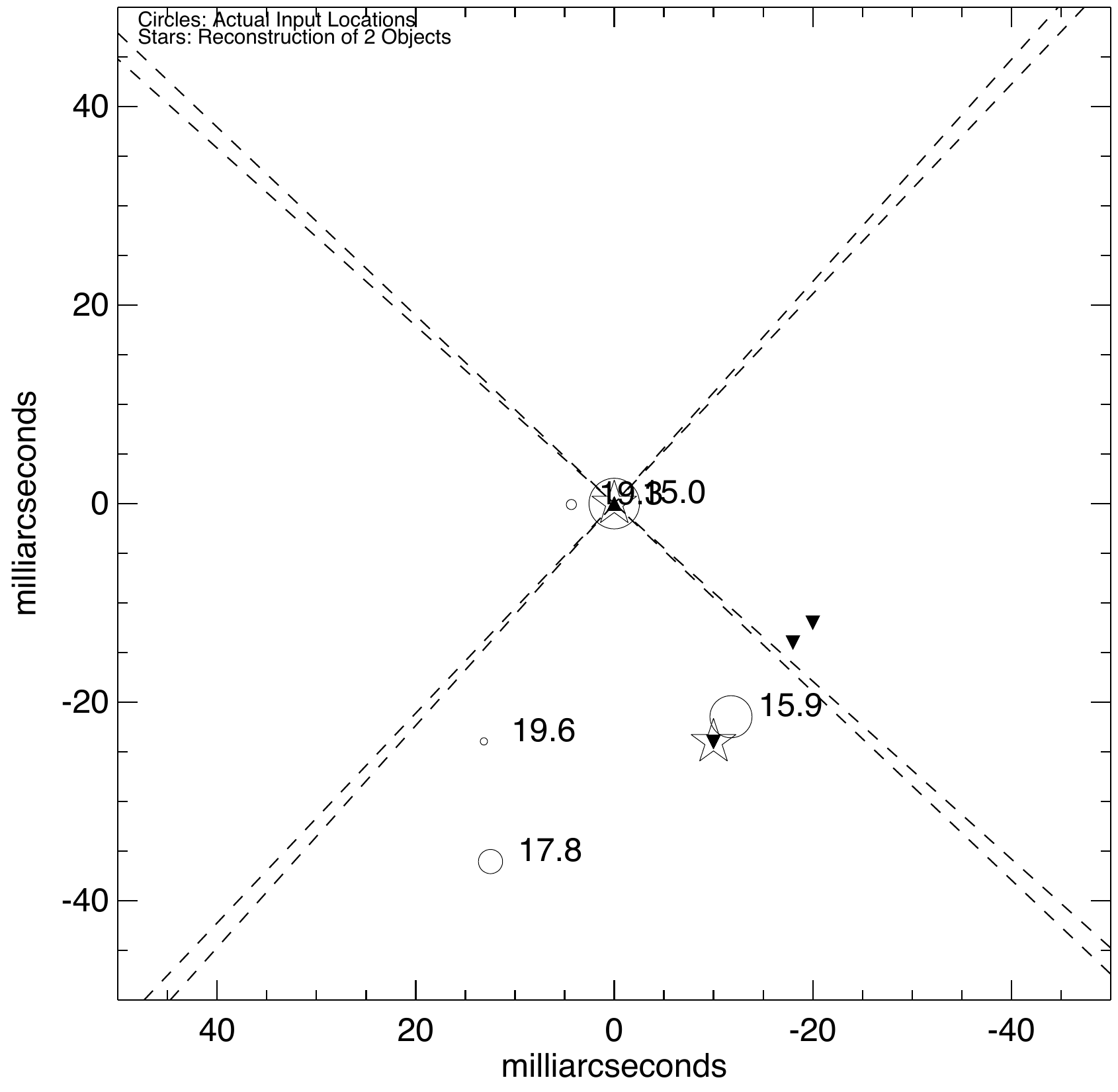}}
  \end{minipage} \hfill
  \begin{minipage}{6.0cm}
\caption{\label{fig:gcim} 
Simulation of the central 50 mas of the Galactic Center.  Here we assume SgrA* has a magnitude of $m_K=15$
and is located at position (0,0).  Several fainter stars, selected
based on a luminosity function extrapolated from current observations, were added at
random positions.  The field was then converted into simulated visibilities, including realistic noise.
We fitted a model to the simulated data to try to recover the
positions of SgrA* and one other source; i.e., our model only included
two objects.  Uncertainties on the fitted positions are estimated
using bootstrapping.  Despite the presence of additional sources in
the field, this simple two-source model reproduces the positions of
the two brightest sources in the input model to within about a milliarcsecond.
Assuming we are able to obtain high signal-to-noise observations,
including more sources in our model should improve the fits further.
\vspace{0.3in}}
\end{minipage}
\end{figure}

With DFPR, ASTRA will achieve sufficient sensitivity to observe stars
in short-period orbits about the Galactic Center.  Bringing the
angular resolution of the Keck Interferometer to bear on these stars
enables a number of interesting science cases.

The simplest is a test for binarity among the population of stars
orbiting the galactic center.  The KI resolution can constrain
binaries with semi-major axes larger than about 10 AU.  These
measurements are important for understanding the origin of the young
stars (i.e., the B stars comprising the S cluster) in the Galactic Center.

One of the main science goals of ASTRA is to decrease the confusion of
stars in the central regions of the GC, and thereby allow higher
precision astrometry. 
The inner 50 mas around SgrA* fits into a single AO
resolution element, and hence the multiple sources here can not be
distinguished with current LGS-AO observations.  By imaging this
50-mas-diameter field centered on
SgrA* with ASTRA, which provides a resolution element approximately 10 times
smaller than AO imaging, we aim to determine the positions of SgrA*
and nearby stars.

Any new stars that we can distinguish would have smaller semi-major
axes than any of the currently known stars.  Assuming a 10 mas
semi-major axis, the orbital period would be 0.1 years.  It would thus
be much simpler to monitor a number of orbits for such stars and to
search for precession of perihelia associated with post-Newtonian
effects (these are discussed in more detail in the following section).
Indeed, simulations have shown that such short-period orbits are 
optimal for measuring post-Newtonian effects, both because their short
periods allow multiple cycles to be observed and because they pass
through a deeper part of the potential well around SgrA*\cite{WMG05}.

In principal, such observations are possible.  However, with only a
single baseline, our PSF has large sidelobes and
we are sensitive to all emission within the 50 mas field of view of
KI. Moreover, the KI baseline has limited rotation for the
low-declination Galactic Center, and so most of our angular resolution
is along a single position angle.  To verify that we can, in fact,
recover positions of multiple sources within a 50 mas field, we
generated simulated observations and fitted multiple-source models to
them.

Using a luminosity function extrapolated from the currently known
stars in the Galactic Center region \cite{WMG05}, we estimate the
number of stars of a given magnitude within a 50 mas region centered
on SgrA*.  We typically find that in addition to SgrA*, this field
contains one $m_K \sim 16$ star, one $m_K \sim 18$ star, and 
stars fainter than $m_K = 19$.  We simulate observations of
this field, including photon noise and realistic astrometric noise.
Finally, we fit the simulated data with a two-component source model
where the positions and fluxes of two stars are free parameters.  

Our simulations show that we can recover the position of a $m_K=16$
star to a few mas or less, even in the presence of SgrA* and two
fainter stars within the 50 mas field of view of KI. Most of the
uncertainty is in the direction orthogonal to the position angle of
the KI baseline.  However, even with 4$^{\circ}$ of baseline rotation
(which is the maximum we can obtain for observations of the Galactic
Center), we can constrain some 2D information.  With higher
signal-to-noise, our constraints get tighter.  Note, too these results
are for a simple two-component model.  More
sophisticated fitting routines, including more than just two
components, will likely improve the accuracy with which we can
recover source positions.

These observations, while providing astrometric information, do not
explicitly require the precise baseline knowledge and monitoring of normal,
dual-star astrometric observations (described in the following
section).   Here stars in a common field are
referenced with respect to one another.  Thus, these ``imaging''
observations are, in many ways, simpler than normal astrometry.

\section{ASTROMETRIC MONITORING OF STARS WITH MULTI-YEAR ORBITS ABOUT THE GALACTIC CENTER}
\label{sec:gcastrom}

\begin{figure}
  \begin{center}
  \includegraphics[scale=0.6,clip=true]{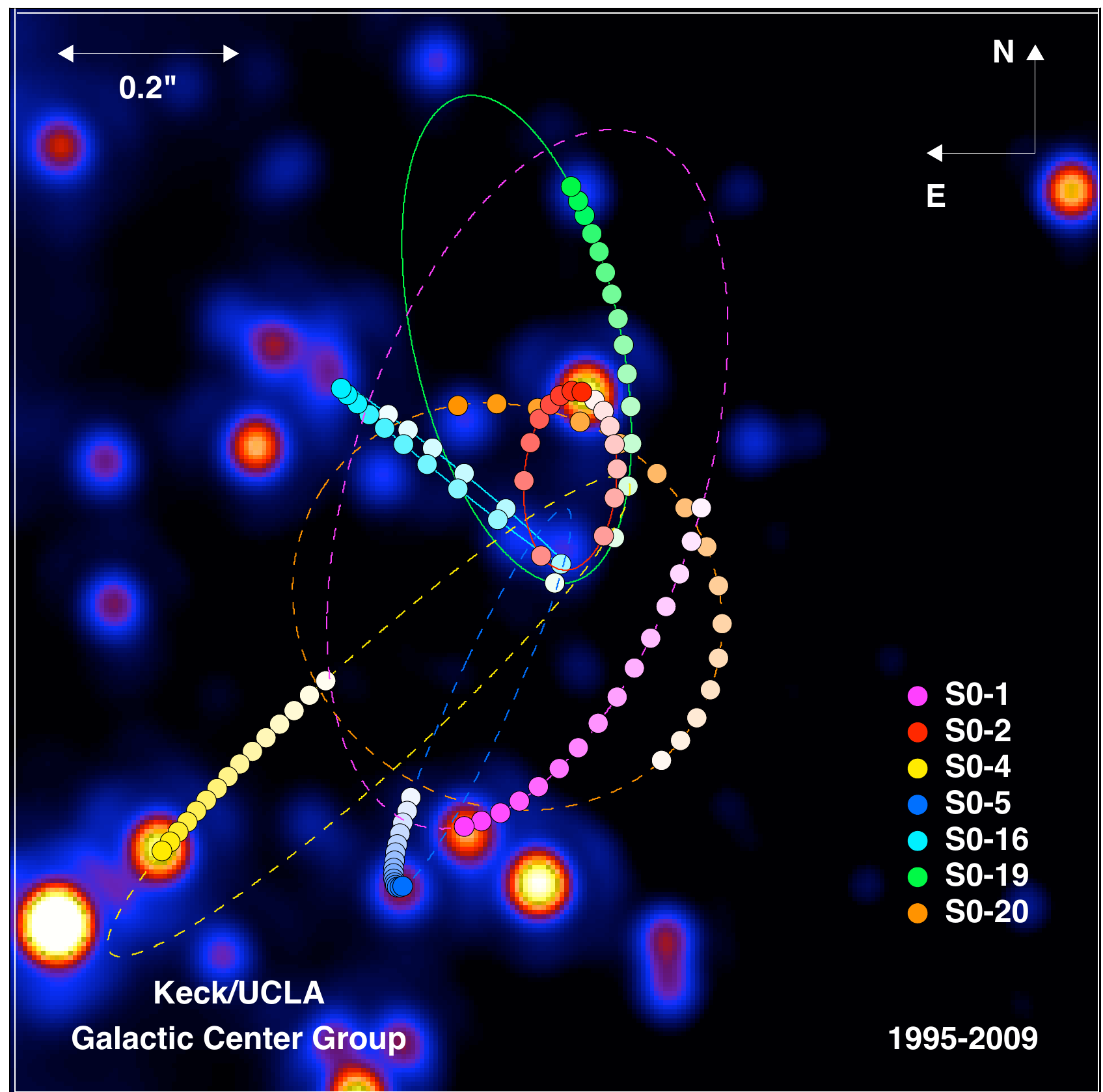}
   \end{center}
   \caption{ \label{fig:gc} 
A plot of the orbits of stars in the Galactic Center determined by NGS- and LGS-AO observations
\cite{GHEZ+08}.  With KI/ASTRA, we will use the enhanced centroiding and deeper confusion limit
afforded by higher angular resolution to improve
the accuracy with which these orbits are measured.
}
  \end{figure}

The orbits of stars within 0$\rlap{.}''4$ of SgrA* have been measured
with LGS-AO observations with an accuracy approaching 150 $\mu$as
(e.g., Figure \ref{fig:gc}).  However, these precisions are only
possible far from SgrA*.  Within the central $0\rlap{.}''1$ or so,
confusion limits the measurement accuracy to 0.5 mas or more.
For example, the star S0-2, which has a $K$ magnitude of 14,
experiences significant confusion upon closest approach
\cite{GILLESSEN+09b}.  This strongly suggests the presence of a star
with $m_K < 16$ within 50 mas of SgrA*.  As discussed above, SgrA*
itself is most likely  confused with this source, as well as with
other faint objects in the central 50 mas.

ASTRA aims to provide $\sim 30$ $\mu$as astrometric accuracy for stars
orbiting around SgrA* \cite{POTT+08}.  ASTRA observations will also be affected by
confusion in the inner regions, although to a lesser extent given the
higher angular resolution.   As for the SgrA* imaging described above,
when stars enter confused regions, we will combine our astrometric
information with imaging information.  The astrometry gives the
position on the sky of the phase center of our interferometric field
of view, while the imaging information provides the position of the
star of interest (and potential confusing sources) relative to this
phase center position.

The anticipated measurement accuracy of
ASTRA is sufficient to constrain post-Newtonian effects, including
prograde general relativistic precession and retrograde precession due
to the extended (dark) mass distribution near to SgrA* \cite{WMG05}.
GR effects can be measured even for single orbits of known stars
(e.g., S0-2) if we have an astrometric precision of $< 100$ $\mu$as.  

In addition to detecting GR effects on stellar orbits in the GC, ASTRA
can measure perturbations to orbits due to the extended matter
distribution around the BH, which may consist of a dark matter cusp
remaining from the Galaxy  formation process.  Moreover, scattering of
stars by encounters with intermediate mass BHs, which may fall in to
the GC under the influence of dynamical friction, would also be
detectable with the KI with an astrometric precision of $\sim 100$ $\mu$as.
Such precision astrometry will also enable measurement of
the BH mass and GC distance  with an unprecedented 1\% 
accuracy, crucial for understanding Galactic structure. 
 
Another topic that can be addressed with the instrumentation
upgrades proposed here is the origin of the observed flares from the
GC \cite{GENZEL+03,GHEZ+04}.  With an astrometric
accuracy of tens of $\mu$as, we can measure the position of the flares
relative to other stars in the GC and thereby determine whether this emission arises
in an accretion disk at several Schwarzschild radii or in an
outflowing jet.   If the flares arise in a disk, future astrometric
observations with $\mu$as precision have the potential to observe centroid
shifts as flare material executes very tight orbits around the BH
\cite{BL05}.

$ $ \\
\centerline{\bf{ACKNOWLEDGMENTS} }
The ASTRA project is made possible by a Major Research Instrumentation (MRI) grant from the National 
Science Foundation (NSF; award \#0619965), and the infrastructure support from the National Aeronautics and 
Space Administration (NASA) to the Keck Interferometer (KI) project, through the Jet Propulsion Laboratory 
(JPL) and the NASA Exoplanet Science Institute (NExScI).

\bibliographystyle{spiebib}   
\bibliography{jae_ref}   

\end{document}